\documentclass[10pt,conference]{IEEEtran}
\IEEEoverridecommandlockouts
\usepackage{cite}
\usepackage{amsmath,amssymb,amsfonts}
\usepackage{algorithmic}
\usepackage{graphicx}
\usepackage{textcomp}
\usepackage{makecell}
\usepackage{xcolor}
\usepackage{stfloats}
\usepackage{multirow}
\usepackage{xcolor}
\usepackage{enumitem}
\usepackage{amsmath,amssymb}

\newcommand{\ie}[0]{i.e.,}
\newcommand{\eg}[0]{e.g.,}
\newcommand{\so}[0]{Stack Overflow}

\usepackage{hyperref} 

\usepackage{fontawesome} 

\usepackage{float}
\usepackage{threeparttable}
\usepackage[most]{tcolorbox}
\usepackage{makecell}
\usepackage{bbding}
\usepackage{amssymb}
\usepackage{pifont}
\usepackage{booktabs}
\usepackage{tikz}
\usepackage{balance}
\usepackage{flushend}

\DeclareRobustCommand*{\IEEEauthorrefmark}[1]{%
  \raisebox{0pt}[0pt][0pt]{\textsuperscript{\footnotesize #1}}%
}

\def\BibTeX{{\rm B\kern-.05em{\sc i\kern-.025em b}\kern-.08em
    T\kern-.1667em\lower.7ex\hbox{E}\kern-.125emX}}

\newcolumntype{C}[1]{>{\centering\arraybackslash}p{#1}}

\newtcbtheorem{Summary}{\bfseries Summary}{enhanced,drop shadow={black!50!white},
  coltitle=black,
  top=0.3in,
  attach boxed title to top left=
  {xshift=1.5em,yshift=-\tcboxedtitleheight/2},
  boxed title style={size=small,colback=pink}
}{summary}

\usepackage{amsmath}
\usepackage{tcolorbox}

\begin{document}

\title{
Aspect-Based API Review Classification: How Far Can Pre-Trained Transformer Model Go?
}



\author{
    \IEEEauthorblockN{Chengran Yang\IEEEauthorrefmark{1}, Bowen Xu\IEEEauthorrefmark{1}\textsuperscript{$\ast$}\thanks{$\ast$ Corresponding author.}, Junaed Younus Khan\IEEEauthorrefmark{2}, Gias Uddin\IEEEauthorrefmark{2}, Donggyun Han\IEEEauthorrefmark{1}, Zhou Yang\IEEEauthorrefmark{1}, David Lo\IEEEauthorrefmark{1}}
    \IEEEauthorblockA{\IEEEauthorrefmark{1} School of Computing and Information System, Singapore Management University}
    \IEEEauthorblockA{\IEEEauthorrefmark{2} Department of Electrical and Computer Engineering, University of Calgary, Alberta, Canada}
    \IEEEauthorblockA{\{cryang, bowenxu.2017, dhan, zyang, davidlo\}@smu.edu.sg, \{junaedyounus.khan, gias.uddin\}@ucalgary.ca}
}


\maketitle

\begin{abstract}
APIs (Application Programming Interfaces) are reusable software libraries and are building blocks for modern rapid software development. Previous research shows that programmers frequently share and search for reviews of APIs on the mainstream software question and answer (Q\&A) platforms like Stack Overflow, which motivates researchers to design tasks and approaches related to process API reviews automatically.
Among these tasks, classifying API reviews into different aspects (e.g., performance or security), which is called the {\em aspect-based API review classification}, is of great importance.
The current state-of-the-art (SOTA) solution to this task is based on the traditional machine learning algorithm. Inspired by the great success achieved by pre-trained models on many software engineering tasks, this study fine-tunes six pre-trained models for the aspect-based API review classification task and compares them with the current SOTA solution on an API review benchmark collected by Uddin et al. The investigated models include four models (BERT, RoBERTa, ALBERT and XLNet) that are pre-trained on natural languages, BERTOverflow that is pre-trained on text corpus extracted from posts on Stack Overflow, and CosSensBERT that is designed for handling imbalanced data.
The results show that all the six fine-tuned models outperform the traditional machine learning-based tool. More specifically, the improvement on the F1-score ranges from $21.0\%$ to $30.2\%$. 
We also find that BERTOverflow, a model pre-trained on the corpus from Stack Overflow, does not show better performance than BERT. The result also suggests that CosSensBERT also does not exhibit better performance than BERT in terms of F1, but it is still worthy of being considered as it achieves better performance on MCC and AUC.

\end{abstract}

\section{Introduction}
Application programming interfaces (APIs) allow data and functionality to be easily shared across different projects. APIs have significantly facilitated the modern software development process.
For example, the Java Software Development Kit comes with thousands of components that developers can conveniently reuse by calling APIs, which reduces both development time and effort~\cite{ahasanuzzaman2018classifying}.
To find the API information that is relevant to their requirements, developers usually use mainstream software question and answer (Q\&A) platforms like \so{}\footnote{\url{https://stackoverflow.com}}, where developers share their reviews on APIs. Such reviews can involve multiple aspects of APIs (e.g., usability and documentation), and understanding the aspects that an API review is about can further aid developers in finding more relevant API information.


\so{} is a popular software question and answer platform with extensive discussion on APIs as well as a rich amount of API reviews~\cite{uddin2021understanding}.
It becomes a valuable resource and attracts many researchers to further leverage the API information for developers
\cite{ahasanuzzaman2018classifying,ahasanuzzaman2020caps}.
Previous research found that API documentation often suffers from quality problems, i.e., incomplete, obsolete, and/or incorrect API documentation~\cite{uddin2015api}.
Thus, many developers seek help and insights about API from other developers in online developer forums (\ie{} \so{})~\cite{uddin2021understanding}.
Uddin et al.~\cite{uddin2021understanding} surveyed 178 software developers to understand their requirements when searching for API information on Q\&A platforms. The results show that although there are plenty of reviews on the same APIs, developers usually value more the API reviews that discuss certain aspects (e.g., is the feature offered by this API scalable?). 
Such facts motivate researchers to design automated approaches that can accurately classify the aspects of API reviews\cite{uddin2017mining,uddin2017automatic,lin2019pattern,uddin2019automatic}.

The task of \emph{aspect-based API review classification} is originally introduced by Uddin et al. in~\cite{uddin2017mining}. The aim of the task is to assign API reviews with pre-defined aspects (\eg{} Usability and Security).
To tackle the task, many approaches have been proposed~\cite{uddin2017automatic,lin2019pattern,uddin2019automatic}.
Uddin et al. propose a machine learning-based approach~\cite{uddin2017automatic} to assign Stack Overflow sentences with pre-defined aspect labels.
Lin et al. propose a pattern-based approach to classify API opinions from Q\&A platforms into aspects~\cite{lin2019pattern}. Furthermore, Uddin et al. propose a state-of-the-art machine learning-based aspect-based API review classification Opiner~\cite{uddin2019automatic}.

In recent years, pre-trained transformer-based models have achieved exceptional performance in many tasks and areas including the
software engineering domain~\cite{zhang2020sentiment,zhou2019lancer,yu2020order,wang2020fret}.
For example, Zhang et al. conduct an empirical study on benchmarking four pre-trained transformer-based models (BERT~\cite{devlin2019bert}, RoBERTa~\cite{liu2019roberta}, ALBERT~\cite{lan2019albert}, and XLNet~\cite{yang2019xlnet}) for sentiment analysis on six software repositories (\eg{} code reviews)~\cite{zhang2020sentiment}.
They find that the pre-trained transformer-based models outperform existing sentiment analysis tools by a big margin.
In the rest of the paper, we refer to the pre-trained transformer-based models collectively as PTMs.

In this work, we focus on the task of \emph{aspect-based API review classification} as well as the corresponding state-of-the-art approach Opiner~\cite{uddin2019automatic}.
Specifically, we would like to investigate the effectiveness of PTMs on this task. This task has the following characteristics:
first, the corresponding state-of-the-art (SOTA) approach is still based on traditional machine learning approaches;
second, the dataset released in~\cite{uddin2019automatic} is based on \so{}, which can be considered as SE-specific data, different from the data that is used to pre-train many of the transformer-based models;
third, the data is imbalanced as the numbers of instances for different aspects of API review vary.
Considering the above, we extend the SOTA work on aspect-based API review classification~\cite{uddin2019automatic} from the following three perspectives:

\begin{enumerate}
    \item \textbf{Standard PTMs vs. SOTA Approach} 
    We compare the performance of the SOTA approach against four well-known PTMs~(\ie{} BERT, RoBERTa, ALBERT, XLNet) for the task. By comparing their performance, we investigate whether PTMs outperform the SOTA approach and by how much margin.
    \item \textbf{Standard BERT vs. Domain-specific BERT} Previous works~\cite{gururangan2020don,tabassum2020code} show that PTMs pre-trained on domain-specific data are expected to outperform a standard PTM that is trained on general data. On the other hand, standard PTMs are trained on the huge amount of data which is usually significantly larger than domain-specific data. In this study, we investigate the effectiveness of a domain-specific variant of BERT named BERTOverflow, which is a BERT-based PTM trained on the data from \so{}~\cite{tabassum2020code}.
    \item \textbf{Standard BERT vs. BERT for imbalanced data} Similarly, we apply a variant of BERT named CostSensBERT which is designed to handle imbalanced data~\cite{madabushi2019cost}. To the best of our knowledge, our work is the first to investigate whether CostSensBERT can contribute to the downstream task in the software engineering domain. 
\end{enumerate}

Overall, we replicate the SOTA approach and adapt six different PTMs to answer the above questions, respectively.

The contributions of this work are as follows:

\begin{enumerate}
    \item We adapt four well-known PTMs for aspect-based API review classification task. Our experiment results show that all the PTMs outperform the SOTA approach by a large margin (ranges from 28.9\% to 30.2\%) in terms of F1.
    In addition, XLNet, RoBERTa, and BERT outperform Opiner in terms of MCC~(ranges from 8.8\% to 27.0\%) and AUC~(ranges from 1.5\% to 6.9\%).
    Exceptionally, ALBERT performs better than Opiner in terms of MCC (by 0.8\%) but worse in terms of AUC (by 4.1\%).
    Overall, our experimental results suggest that the PTMs should be considered as baselines for the following works.
    \item By comparing the performance of the standard BERT with BERTOverflow, we find that the standard BERT slightly outperforms BERTOverflow in terms of F1 (by 3.9\%). However, BERTOverflow performs worth than BERT in terms of MCC and AUC by 77.9\% and 28.8\%, respectively.
    We suggest that BERT may be better to be used for many Stack Overflow related tasks instead of BERTOverflow. The original BERTOverflow paper has only investigated its effectiveness on the SE-specific name entity recognition task.
    This being said, BERTOverflow still outperforms Opiner by a large margin (by 25.3\%) in terms of F1. However, BERTOverflow perform worse than Opiner in terms of MCC and AUC by 72.0\% and 23.8\%, respectively.
    \item By comparing BERT and CostSensBERT, we find that BERT performs better than CostSensBERT in terms of F1 (by 7.5\%) but worse on MCC and AUC (by 8.9\% and 3.7\%, respectively).
    Considering that CostSensBERT achieves better performance in terms of MCC and AUC than BERT, it indicates that CostSensBERT may still be worthy of being considered as baselines in future works.
    Also, CostSensBERT outperforms Opiner on F1, MCC, and AUC by a large margin (21.0\%, 38.3\%, 10.9\%, respectively).
\end{enumerate}

\section{Background}
In this section, we introduce the problem formulation of aspect-based API review classification. 
In addition, we describe the state-of-the-art approach Opiner. Finally, we introduce the four popular PTMs (BERT, RoBERTa, XLNet, and ALBERT) and two variants (BERTOverflow and CostSensBERT).

\begin{table*}[!htbp]
    \centering
    \caption{The Definition of API Review Aspects}

\resizebox{\linewidth}{!}{\begin{tabular}{lll}
\toprule
\textbf{Aspect}        & \textbf{Definition}                 
& \textbf{Examples} \\
\midrule
\multirow{2}{*}{Performance}    & How the software performs in terms of speed or other
&\multirow{2}{*}{The object conversion in GSON is fast} \\
                                &  performance issues? &                \\\hline
 \multirow{2}{*}{Usability}    & Is the software easy to use? How well is the software
&\multirow{2}{*}{GSON is easy to use}\\
                                & designed to meet specific development requirements?&\\
\hline
\multirow{2}{*}{Security}     & Does the usage of the software pose any security&
The network communication using the HTTPClient API is\\
                                & threat?& not secure\\\hline
\multirow{1}{*}{Bug}        & The opinion is about a bug related to the software. &
\multirow{1}{*}{GSON crashed when converting large JSON objects}\\
\hline
\multirow{2}{*}{Community}    & How supportive/active the community
&\multirow{2}{*}{The GSON mailing list is active}\\
                                & (e.g., mailing list) related to the software?&\\\hline
\multirow{3}{*}{Compatibility}  & Whether the usage of the software require of another
&\multirow{3}{*}{Spring uses Jackson to provide JSON parsing}\\
                                & software or the underlying development/deployment&\\
                                & environment. &\\
\hline
\multirow{2}{*}{Documentation}  & Documentation about the software is available and is 
&\multirow{2}{*}{GSON has good documentation}\\
                                & of good/bad quality.&\\\hline
\multirow{2}{*}{Legal}  & The usage of the software does/does not require any  
&\multirow{2}{*}{GSON has an open-source license}\\
                                & legal considerations.&\\\hline
\multirow{2}{*}{Portability}  & The opinion about the usage of the software across 
&\multirow{2}{*}{GSON can be used in windows, linux and mobile}\\
                                &different platforms.&\\\hline
\multirow{2}{*}{OnlySentiment}  & Opinions about the software without specifying any 
&\multirow{2}{*}{I like GSON}\\
                                &  particular aspect/feature of the software.&\\
\hline
\multirow{2}{*}{Others}  & The opinion about the aspect cannot be labelled using 
&\multirow{2}{*}{What is the difference between swing and awt}\\
                                & any of the above categories.&\\
\hline
                                                           \end{tabular}}
    \label{tab:definitionofaspect}
    \vspace{-3mm}
\end{table*}

\subsection{Problem Formulation}\label{sec:formulation}


In~\cite{uddin2017mining}, Uddin et al. surveyed 178 software developers and summarized 11 API review aspects developers prefer to see in the \so{}.
The definition of each aspect is shown in Table~\ref{tab:definitionofaspect} as well as the corresponding example instances extracted from \so{}.

Given API reviews in the form of sentences, the goal of aspect-based API review classification is to assign each sentence with one or multiple pre-defined aspects.
By following the SOTA approach Opiner \cite{uddin2019automatic}, this task is formulated as multiple binary classification problems while each problem corresponds to predicting if a sentence (in an API review) refers to a particular aspect (e.g., Performance, Usability, Legal).
In other words, each aspect corresponds to a binary classification problem. The input is a given API review while the output is the prediction if the review is related to the aspect.  

\subsection{Opiner}\label{sec:opiner}
Opiner is a tool
developed by Uddin et al. which aims to leverage the API-related information from \so{} for helping developers~\cite{uddin2017mining,uddin2019automatic}.
One of the key functionalities of Opiner is API review classification.
For each API review aspect, Opiner performs two main steps: feature extraction and classification.
In the feature extraction step, Opiner tokenizes and vectorizes the data instances (\ie{} sentences) into n-grams.
In particular, Opiner uses n = 1,2,3 for n-grams, \ie{} unigrams (one single word) to trigrams (a sequence of three words).
Then, Opiner normalizes the n-grams by applying a standard TF-IDF algorithm that converts the collection of instances to the set of matrix of TF-IDF features.
In the classification step, two traditional machine learning models are selected as candidate classifiers: SVM and Logistic Regression.
To evaluate Opiner, Uddin et al. apply 10-fold cross-validation and pick the best performing classifier in every aspect.

\subsection{Pre-trained Transformer-based Models}
\label{sec:ptms}

\begin{itemize}
    \item \noindent\textbf{BERT} (\textbf{B}idirectional \textbf{E}ncoder \textbf{R}epresentations from
\textbf{T}ransformers) is a language representation model which proposed by Google~\cite{devlin2019bert}.
BERT is conceptually simple and empirically powerful~\cite{zhang2020sentiment}.
Its key technical novelty is applying the bidirectional training of Transformer, a popular attention model for language modeling.
BERT is pre-trained by two unsupervised tasks, Masked Language Modeling (MLM) and Next Sentence Prediction (NSP).
In MLM, some words from input sentences are randomly masked and BERT attempts to predict the masked words, based on the context provided by the other, non-masked, words in the sequence.
In NSP, BERT predicts if one sentence follows another.
BERT has achieved outstanding performance for many software tasks(\eg{} sentiment classification for software artifacts~\cite{zhang2020sentiment}).
\item \textbf{RoBERTa} (\textbf{R}obustly \textbf{O}ptimized \textbf{BERT} \textbf{A}pproach) is a modified version of BERT on the training procedure~\cite{liu2019roberta}.
The modification includes: (1) training the model longer time
with bigger batches, over more data; (2) removing the NSP task; (3) training on longer sequences; and (4) dynamically changing the masking pattern applied to the training data in the MLM task. Yang et al. \cite{yang2021deepscc} fine-tuned RoBERTa to classify the programming language type of the source code.
\item \textbf{ALBERT} (\textbf{A} \textbf{L}ite \textbf{BERT}) is a lite version of BERT which scales much better compare to the original BERT~\cite{lan2019albert}.
The technical design is motivated when model increases become harder due to GPU/TPU memory limitations and longer training times.
To address these problems, ALBERT utilizes two parameter reduction techniques, factorized embedding parameterization and cross-layer
parameter sharing, to lower memory consumption and increase the training speed of BERT.

\item \textbf{XLNet} is a pre-trained model that uses a permutation language modeling objective to leverage the strengths of autoregressive (AR) and autoencoding (AE) methods while avoiding their limitations~\cite{yang2019xlnet}.
XLNet is capable to learning bidirectional contexts by maximizing the expected likelihood over all permutations of the factorization order.
Aside from using permutation language modeling, XLNet utilizes Transformer XL~\cite{dai2019transformer}, which improves its performance further.
The key technical novelties of Transformer XL are segment recurrence mechanism and relative encoding scheme.

\item \textbf{BERTOverflow} is a domain-specific version of BERT which is trained based on 152 million sentences from \so{}~\cite{tabassum2020code}.
It is originally applied for the task of software-specific name entity recognition based on \so{} data.
Similarly, we also focus on \so{} data in this study.
Thus, BERTOverflow could potentially perform well for our target task, \ie{} aspect-based API review classification.

\item \textbf{CostSensBERT}\label{sec:costsensbert}
is also a modified version of BERT which is designed to handle imbalanced data by incorporating cost-sensitivity~\cite{madabushi2019cost}.
The main idea of CostSensBERT is that increasing the cost of predicting the classification of an ``important'' class wrong and corresponding decrease the cost of predicting a less important class wrong.
Considering the used data in this study is imbalanced, intuitively, CostSensBERT could be a suitable variant for our task.
\end{itemize}

\section{Methodology}
This section first describes the dataset used in this work. Then, we elaborate on the implementation details of all approaches. Lastly, we describe the evaluation metrics and experimental settings.

\subsection{Dataset}\label{sec:datasets}
To facilitate the comparison, we use the same dataset constructed by Uddin et al.~\cite{uddin2019automatic}. The dataset consists of 4,522 sentences extracted from 1,338 posts that discussed APIs in \so{}. Each sentence is manually labeled as one or multiple API review aspects~\cite{uddin2019understanding}.
Among all the sentences in the dataset, 4,307 of them are labeled by only one API aspect, 209 sentences are labeled with two aspects, and 6 sentences are assigned with three or more aspects.

\begin{table}[]
    \caption{Distribution of Aspects in the Dataset
    }
    \centering
    \begin{tabular}{ll}
    \toprule
    \textbf{Aspect} & \textbf{\# (\%) of Instances}\\
    \midrule
        Performance& 348 (7.7) \\
        Usability &1,437 (31.8)\\
        Security &163 (3.6)\\
        Bug&189 (4.2)\\
        Community&93 (2.1)\\
        Compatibility&93 (2.1)\\
        Documentation&256 (5.6)\\
        Legal&50 (1.1)\\
        Portability&70 (1.5)\\
        OnlySentiment&348 (7.7)\\
        Others&1,699 (37.6)\\\hline
        Total &4,522\\
    \bottomrule

    \end{tabular}
    \label{tab:percentage}
    \vspace{-3mm}
\end{table}

The data distribution shown in Table \ref{tab:percentage} indicates that the dataset is imbalanced, \ie{} the numbers of instances differ across the aspects.
Among all the instances, the ratio of every API review aspect varies from 1.1\% to 37.6\% in this dataset. Two API review aspects contain more than 30\% instances. In comparison, 6 out of 11 aspects correspond to less than 5\% samples.
\subsection{Sample Strategy}\label{sec:sampling}


As mentioned in Section~\ref{sec:datasets}, the dataset is imbalanced. In \cite{uddin2019automatic}, Uddin et al. utilized two different sampling strategies to mitigate the data imbalance issue: balanced sample strategy with undersampling; imbalanced sample strategy without undersampling.
We apply both sampling strategies on all approaches.

\noindent\textbf{With Undersampling} For each given API review aspect, we regard all the corresponding instances as positive samples and randomly select the same number of instances from remaining aspects (labeled as other aspects) as negative samples.

\noindent\textbf{Without Undersampling} For each given API review aspect, we regard all the corresponding instances as positive samples, and all the remaining instances as negative samples.

\subsection{Experimental Setting}
In this study, we follow the same experimental setting in Opiner~\cite{uddin2019automatic} and apply the setting to all the approaches for a fair comparison. 
We employ a standard stratified 10-fold cross-validation to evaluate the approaches.
Moreover, for the pre-trained transformer-based models, we split the data into ten folds, while eight for fine-tuning, one for validation, and one for test.
In adition, we used AdamW as the optimizer for all the PTMs which is widely-used in many works, e.g.,~\cite{loshchilov2017decoupled,lan2019albert,liu2019roberta}.
In adition, we utilize AdamW which is widely-used in many works~(e.g.,~\cite{loshchilov2017decoupled,lan2019albert,liu2019roberta}) as the optimizer for all the PTMs.

\subsection{Implementations} \label{sec:implementations}
\textbf{The SOTA approach: Opiner}
Since the replication package of Opiner has not been released online yet and the main algorithms used in Opiner are simple, we re-implemented it based on the description from the original paper and consulted with the corresponding authors to confirm a few unclear details (e.g., the selection of optimal hyper-parameters).
We managed to replicate Opiner and observed even slightly better performance than the results reported in the original paper by the above means.
We further discuss the potential threat in Section~\ref{sec:threats}.

\textbf{Pre-trained Transformer-based Approaches}
In this study, we consider four popular and state-of-the-art pre-trained transformer-based models which have been utilized in many other software tasks~\cite{biswas2020achieving,zhang2020sentiment,khan2021automatic}, including BERT, RoBERTa, ALBERT, XLNet. We also apply two PTM variants: BERTOverflow~\cite{tabassum2020code} that is pre-trained with software engineer in-domain data; CostSensBERT~\cite{madabushi2019cost} that designed to handle imbalanced data.
To adapt the four well-known PTMs for our task, we follow a standard way by adding a dropout layer and a linear layer~\cite{devlin2019bert} on the top of the PTMs.
We fine-tune the PTMs by feeding the training data mentioned in Section~\ref{sec:datasets}.

Inspired by the hyper-parameter setting of BERT, we tune three key hyper-parameters with the same scope mentioned in the original paper~\cite{devlin2019bert}, they are (1). {\em learning rate}; (2) {\em batch size}; (3) {\em number of epochs}.
We tune the learning rate by varying three different values, 5e-5, 3e-5, and 1e-5 while varying 16 and 32 for Batch size.
We set the maximum number of the epoch as 5. 

In addition, to adapt CostSensBERT, one additional hyper-parameter named {\em Class Weight} needs to be set, which refers to the cost increasing for obtaining the low-frequency class label wrong. We set its value as (1,20), which is suggested in the paper~\cite{madabushi2019cost}.

We implement all PTMs except CostSensBERT by using a popular deep learning library Hugging Face Transformer\footnote{\url{https://huggingface.com}}. For CostSensBERT, we reuse its replication package\footnote{\url{https://github.com/H-TayyarMadabushi/Cost-Sensitive\_Bert\_and\_Transformers}}.
BERT, RoBERTA, XLNet, and ALBERT released several versions with different parameter sizes. We use the same model version by following their original papers for all PTM approaches. The versions of the PTMs presented in Table \ref{tab:modelparameter} are the corresponding names in the Hugging Face Transformer library.

\begin{table}[h]
\vspace{-3mm}
    \centering
    \caption{Model Version of Considered PTMs}
    \begin{tabular}{c|c}
    \toprule
   \textbf{Model Architecture} &\textbf{Model Version}\\
    \midrule
    BERT & bert-base-uncased\\
    ALBERT & albert-base-v2\\
    RoBERTa & roberta-base\\
    XLNet & xlnet-base-cased\\
    BERTOverflow & jeniya/BERTOverﬂow\\
    CostSensBERT & bert-base-uncased\\
    \bottomrule
    \end{tabular}
    \label{tab:modelparameter}
    \vspace{-3mm}
\end{table}

\subsection{Evaluation Metrics}\label{sec:eval_metrics}

By following the latest work~\cite{uddin2019automatic}, we use the same five evaluation metrics to measure all the approaches. They are weighted precision~(denoted as P), weighted recall~(denoted as R), weighted F1~(denoted as F1),
the Matthews Correlation Coefficient score (MCC), and the score of Weighted Area Under ROC Curve~(AUC).
Same as~\cite{uddin2019automatic}, we also regard F1 as the main evaluation metric for a fair comparison.
The formulas of P, R, and F1 are calculated as follows:

\begin{equation}\label{p}
\begin{split}
Pre_{category}=\frac{\#TP_{category}}{\#TP_{category}+\#FP_{category}}
\end{split}
\end{equation}
\begin{equation}\label{R}
\begin{split}
Rec_{category}=\frac{\#TP_{category}}{\#TP_{category}+\#FN_{category}}
\end{split}
\end{equation}
\begin{equation}\label{f1}
\begin{split}
F1_{category}=2\times\frac{Pre_{category}\times R_{category}}{Pre_{category}+R_{category}}
\end{split}
\end{equation}

\begin{equation}\label{weighted_P}
\begin{split}
P = \frac{1}{2}\times \sum_{category=1}^{\#category}n_{category}\times Pre_{category}
\end{split}
\end{equation}
\begin{equation}\label{weighted_R}
\begin{split}
R = \frac{1}{2}\times \sum_{category=1}^{\#category}n_{category}\times Rec_{category}
\end{split}
\end{equation}
\begin{equation}\label{weighted_F1}
\begin{split}
F1 = \frac{1}{2}\times \sum_{category=1}^{\#category}n_{category}\times F1_{category}
\end{split}
\end{equation}

As the problem is formulated as binary classification problem, the number of category is set to 2.
For every aspect, $TP_{category}$ refers to the number of true-positive samples of a particular category (\ie{} sentences are correctly classified with the aspect); $FP_{category}$ refers to the number of false-positive samples of a particular category~(\ie{} sentences are mistakenly classified with the aspect); $FN_{category}$ refers to the numbers of false-negative samples of a particular category~(\ie{} sentences mistakenly classified as other aspects).
$n_{category}$ is the number of instances of a particular category.





MCC is a measure of the quality of binary classification, which takes into account true and false positives and negatives. The formula of MCC is calculated as follow:

\begin{small}
\begin{equation}\label{mcc}
\begin{split}
\resizebox{0.9\hsize}{!}{
$MCC=\frac{TP\times TN-FP\times FN}{\sqrt{(TP+FP)(TP+FN)(TN+FP)(TN+FN)}}$
}
\end{split}
\end{equation}
\end{small}

\noindent AUC stands for the area under the Receiver Operating Characteristic Curve (ROC-AUC).
It is threshold independent measure, which give us the insight from ranking prediction perspective.
It represents to the probability that a randomly chosen negative example will have a smaller estimated probability of belonging to the positive class than a randomly chosen positive example~\cite{hand2001simple,huang2005using}. 
The formula of AUC score is calculated as follow:
\begin{equation}\label{mcc}
\begin{split}
AUC = \frac{S_{0}-n_{0}(n_{0}+1)/2}{n_{0}n_{1}}
\end{split}
\end{equation}

Where $n_{0}$ and $n_{1}$ are the numbers of positive and negative samples, respectively, and $S_{0} = \sum r_{i}$, where $r_{i}$ is the rank of the $i^{th}$ positive example in the descending list of output score produced by every model.

\section{Evaluation}
In this section, we answer three research questions:

\noindent\textbf{RQ1}. Can pre-trained transformer-based models achieve better performance than the state-of-the-art approach which is based on traditional machine learning models?

\noindent\textbf{RQ2}. Can domain-specific BERT (\ie{} BERTOverflow) achieve better performance than the standard BERT? If not, can it still outperform Opiner?

\noindent\textbf{RQ3}. Can the BERT for imbalanced data (\ie{} CostSensBERT) achieve better performance than the standard BERT? If not, can it still outperform Opiner?

For each research question, we first describe the corresponding motivation and then report and analyze the performance of seven approaches on aspect-based API review classification task. 
For each API review aspect, we evaluate the performance in terms of five evaluation metrics (i.e., $P, R$, $F1$, $MCC$, and $AUC$) as introduced in Section~\ref{sec:eval_metrics}.
\begin{table}[!htbp]
\vspace{-3mm}
    \centering
    \caption{Average Performance of Approaches on all the aspects}
    \begin{tabular}{c|C{1.6cm}C{1.5cm}C{1.5cm}}
    \toprule
    Model Name & Avg F1  & Avg MCC  & Avg AUC     \\
    \midrule
    Opiner     &  0.724&  0.478 & 0.734\\
    ALBERT & 0.933& 0.482 & 0.704\\
    BERT & 0.942& 0.607 & 0.785\\
    RoBERTa & \textbf{0.943}& 0.585 & 0.775\\
    XLNet & 0.938&0.520&0.745\\
    BERTOverflow & 0.907 & 0.134 & 0.559\\
    CostSensBERT & 0.876 & \textbf{0.661} & \textbf{0.814} \\ 
    \bottomrule
    \multicolumn{4}{c}{The numbers in bold show the highest scores among all approaches.}
    \end{tabular}

    \label{tab:avgf1fortransformer}
\end{table}

\begin{table*}[]
\caption{Detailed Performance of PTMs and Opiner. (\faThumbsOUp: the best performer in terms of F1. The highest F1 score is in \textbf{bold})}
    \label{tab:transformerperformance}
    \centering
    \resizebox{\linewidth}{0.63\linewidth}{%

    \begin{tabular}{C{0.2\linewidth}C{0.15\linewidth}C{0.1\linewidth}C{0.1\linewidth}C{0.1\linewidth}C{0.1\linewidth}C{0.1\linewidth}}
    \toprule
    Aspect         & Approach    & Precision  & Recall  & F1    & MCC   & AUC    \\
    \midrule
\multirow{7}{*}{Performance}      & Opiner    & 0.680     & 0.796  & 0.732 & 0.433 & 0.711 \\\cline{2-7}
                                  & ALBERT        & 0.952     & 0.951  & 0.950 & 0.652 & 0.810 \\
                                  & BERT           & 0.963     & 0.962  & 0.962 & 0.735 & 0.863 \\
                                  & RoBERTa         & 0.966     & 0.965  & 0.965 & 0.761 & 0.875 \\
                                  & XLNet  \faThumbsOUp         & 0.966     & 0.966  & \textbf{0.966} & 0.759 & 0.868 \\\cline{2-7}
                                  & BERTOverflow  & 0.882     & 0.923  & 0.898 & 0.149 & 0.545\\\cline{2-7}
                                  &CostSensBERT  &0.913	&0.878	&0.887	&0.773	&0.878\\\hline
\multirow{7}{*}{Usability}        & Opiner  & 0.649     & 0.780  & 0.707 & 0.371 & 0.680 \\\cline{2-7}
                                  & ALBERT       & 0.783     & 0.779  & 0.779 & 0.496 & 0.748 \\
                                    & BERT           & 0.799     & 0.795  & 0.795 & 0.532 & 0.766 \\
                                    & RoBERTa \faThumbsOUp       & 0.804     & 0.795  & \textbf{0.797} & 0.544 & 0.778 \\
                                    & XLNet           & 0.794     & 0.784  & 0.785 & 0.518 & 0.764 \\\cline{2-7}
                                    & BERTOverflow & 0.747     & 0.742  & 0.742 & 0.412 & 0.707 \\\cline{2-7}
                                    & CostSensBERT &0.777	&0.769	&0.770	&0.547	&0.771 \\\hline
\multirow{7}{*}{Security}         & Opiner  & 0.876     & 0.786  & 0.823 & 0.678 & 0.835 \\\cline{2-7}
                                  & ALBERT         & 0.981     & 0.975  & 0.975 & 0.681 & 0.821 \\
                                 & BERT            & 0.986     & 0.974  & 0.978 & 0.773 & 0.894 \\
                                 & RoBERTa  \faThumbsOUp       & 0.983     & 0.985  & \textbf{0.983} & 0.757 & 0.866 \\
                                 & XLNet           & 0.986     & 0.976  & 0.979 & 0.774 & 0.893 \\\cline{2-7}
                                 & BERTOverflow &0.930	&0.964	&0.947	&0.000	&0.500 \\\cline{2-7}
                                 & CostSensBERT &0.935	&0.921	&0.918 &0.825 &0.921 \\\hline
\multirow{7}{*}{Community}          & Opiner  & 0.641     & 0.534  & 0.572 & 0.242 & 0.615 \\\cline{2-7}
                                  & ALBERT        & 0.962     & 0.980  & 0.970 & 0.049 & 0.512 \\
                                  & BERT            & 0.976     & 0.981  & 0.976 & 0.331 & 0.592 \\
                                  & RoBERTa  \faThumbsOUp       & 0.976     & 0.981  & \textbf{0.976} & 0.352 & 0.624 \\
                                  & XLNet           & 0.972     & 0.980  & 0.974 & 0.269 & 0.591 \\\cline{2-7}
                                  & BERTOverflow & 0.959     & 0.979  & 0.969 & 0.000 & 0.500\\\cline{2-7}
                                  & CostSensBERT &0.933	&0.804&	0.847&	0.423&	0.664 \\\hline
\multirow{7}{*}{Compatibility}    & Opiner  & 0.619     & 0.664  & 0.637 & 0.249 & 0.622 \\\cline{2-7}
                                  & ALBERT        & 0.959     & 0.979  & 0.969 & 0.000 & 0.500 \\
                                  & BERT  \faThumbsOUp          & 0.973     & 0.981  & \textbf{0.975} & 0.299 & 0.587 \\
                                  & RoBERTa         & 0.964     & 0.980  & 0.971 & 0.074 & 0.523 \\
                                  & XLNet           & 0.966     & 0.980  & 0.972 & 0.107 & 0.538 \\\cline{2-7}
                                  & BERTOverflow & 0.960     & 0.980  & 0.970 & 0.000 & 0.500\\\cline{2-7}
                                  & CostSensBERT &0.897	&0.847	&0.869	&0.306	&0.596\\\hline                                    
\multirow{7}{*}{Portability}      & Opiner  & 0.843     & 0.757  & 0.762 & 0.623 & 0.800 \\\cline{2-7}
                                  & ALBERT          & 0.981     & 0.985  & 0.983 & 0.375 & 0.667 \\
                                  & BERT            & 0.991     & 0.990  & 0.990 & 0.693 & 0.840 \\
                                  & RoBERTa  \faThumbsOUp       & 0.992     & 0.992  & \textbf{0.992} & 0.746 & 0.883 \\
                                  & XLNet         & 0.962     & 0.980  & 0.970 & 0.049 & 0.512 \\\cline{2-7}
                                  & BERTOverflow & 0.969     & 0.985  & 0.977 & 0.000 & 0.500 \\\cline{2-7}
                                  & CostSensBERT &0.959	&0.930	&0.933	&0.786	&0.880	\\   \hline                              
\multirow{7}{*}{Documentation}    & Opiner  & 0.729     & 0.826  & 0.772 & 0.527 & 0.759 \\\cline{2-7}
                                  & ALBERT         & 0.950     & 0.955  & 0.948 & 0.482 & 0.685 \\
                                  & BERT  \faThumbsOUp          & 0.964     & 0.964  & \textbf{0.964} & 0.654 & 0.818 \\
                                  & RoBERTa         & 0.960     & 0.961  & 0.960 & 0.616 & 0.794 \\
                                  & XLNet           & 0.962     & 0.961  & 0.961 & 0.632 & 0.813 \\\cline{2-7}
                                  & BERTOverflow & 0.908     & 0.944  & 0.924 & 0.120 & 0.540\\\cline{2-7}
                                  & CostSensBERT &0.919	&0.905	&0.913	&0.764 &0.875\\\hline
\multirow{7}{*}{Bug}              & Opiner  & 0.849     & 0.742  & 0.786 & 0.598 & 0.794 \\\cline{2-7}
                                 & ALBERT         & 0.966     & 0.965  & 0.964 & 0.557 & 0.739 \\
                                 & BERT            & 0.974     & 0.974  & 0.974 & 0.664 & 0.810 \\
                                 & RoBERTa         & 0.962     & 0.980  & 0.970 & 0.549 & 0.712 \\
                                 & XLNet  \faThumbsOUp         & 0.974     & 0.974  & \textbf{0.974} & 0.667 & 0.814 \\\cline{2-7}
                                 & BERTOverflow  & 0.919     & 0.959  & 0.938 & 0.000 & 0.500\\\cline{2-7}
                                 &CostSensBERT  &0.905	&0.876	&0.889	&0.739 &0.856	\\\hline
\multirow{7}{*}{Legal}            & Opiner  & 0.877     & 0.820  & 0.835 & 0.640 & 0.810 \\\cline{2-7}
                                  & ALBERT         & 0.983     & 0.990  & 0.986 & 0.526 & 0.709 \\
                                  & BERT \faThumbsOUp           & 0.996     & 0.995  & \textbf{0.995} & 0.788 & 0.889 \\
                                  & RoBERTa         & 0.996     & 0.995  & 0.995 & 0.784 & 0.869 \\
                                  & XLNet           & 0.995     & 0.995  & 0.995 & 0.753 & 0.849 \\\cline{2-7}
                                  & BERTOverflow & 0.978     & 0.989  & 0.983 & 0.000 & 0.500\\\cline{2-7}
                                  &CostSensBERT &0.930	&0.911	&0.916	&0.789	&0.881\\\hline
\multirow{7}{*}{OnlySentiment}    & Opiner   & 0.814     & 0.702  & 0.750 & 0.546 & 0.769 \\\cline{2-7}
                                  & ALBERT          & 0.947     & 0.951  & 0.948 & 0.620 & 0.774 \\
                                  & BERT           & 0.946     & 0.948  & 0.947 & 0.612 & 0.784 \\
                                  & RoBERTa \faThumbsOUp        & 0.951     & 0.952  & \textbf{0.951} & 0.647 & 0.801 \\
                                  & XLNet           & 0.948     & 0.951  & 0.948 & 0.624 & 0.780 \\\cline{2-7}
                                  & BERTOverflow & 0.917     & 0.927  & 0.920 & 0.402 & 0.670\\\cline{2-7}
                                  & CostSensBERT &0.912	&0.908	&0.910	&0.679	&0.828\\\hline
\multirow{7}{*}{General Features} & Opiner  & 0.694     & 0.619  & 0.653 & 0.351 & 0.674 \\\cline{2-7}
                                 & ALBERT         & 0.798     & 0.789  & 0.789 & 0.563 & 0.782 \\
                                 & BERT            & 0.812     & 0.808  & 0.808 & 0.594 & 0.792 \\
                                 & RoBERTa  \faThumbsOUp       & 0.814     & 0.812  & \textbf{0.812} & 0.601 & 0.798 \\
                                 & XLNet           & 0.798     & 0.797  & 0.795 & 0.565 & 0.775 \\\cline{2-7}
                                 & BERTOverflow & 0.718     & 0.716  & 0.716 & 0.388 & 0.684\\\cline{2-7}
                                 & CostSensBERT &0.797	&0.794	&0.793	&0.611	&0.804\\
\bottomrule       
    \end{tabular}}

\end{table*}

\noindent\textbf{RQ1} {Can pre-trained transformer-based models achieve better performance than the state-of-the-art approach which is based on traditional machine learning models?}

\noindent \textbf{Motivation}
Previous studies have shown the great potential of pre-trained transformer-based models on many software engineering tasks, \eg{} sentiment analysis for software data~\cite{zhang2020sentiment} and code summarization~\cite{wang2020fret}.
However, the efficacy of the pre-trained transformer-based models for various types of software data still remains unclear. Hence, we investigate the effectiveness of four popular state-of-the-art PTMs (\ie{} BERT, RoBERTa, ALBERT, XLNet) for the task of aspect-based API review classification.

\noindent \textbf{Results \& Analysis}
To answer RQ1, we compare the performance of four popular PTMs (\ie{} BERT, RoBERTa, ALBERT, XLNet) with the state-of-the-art approach (\ie{} Opiner).
We present the summative and the detailed experimental results of the approaches in Table \ref{tab:avgf1fortransformer} and Table \ref{tab:transformerperformance}, respectively.
For the summative result in Table~\ref{tab:avgf1fortransformer}, we calculate the arithmetic average of the used evaluation metrics of each approach across all the aspects as avg. F1, avg. MCC, and avg. AUC, respectively. 
From Table \ref{tab:avgf1fortransformer}, we found that all the four PTMs outperform Opiner in terms of F1 while RoBERTa is the best performer. The improvement ranges from 28.9\% to 30.2\%.


We also found that the best performer for all the aspects is not always the same. In particular, among all the 11 aspects, RoBERTa achieves the best performance for six aspects (i.e., Usability, Security, Community, Portability, OnlySentiment, General Features), BERT is the best for three aspects (i.e., Compatibility, Documentation, Legal), and XLNet is the best for two aspects, (i.e., Performance, Bug).

In addition, we observed that most of the PTMs outperform Opiner on all the evaluation metrics but there is an exception.
ALBERT performs worse than Opiner in terms of average AUC score by 4.08\%.
More specifically, the MCC and AUC of ALBERT are worse than Opiner in the three aspects (i.e., Community, Compatibility, Legal) by at least 60\%. And all three aspects have less than 3\% instances in the whole dataset.
ALBERT even has an MCC score of 0 on the Compatibility aspect.
Conversely, the other three PTMs outperform Opiner on two or all aspects in terms of MCC score.
One potential reason is that: 
Considering that ALBERT is a lightweight version of BERT, the parameter reduction techniques utilized in ALBERT may result in potential performance degradation.

Moreover, comparing the best PTM performer (\ie{} RoBERTa) with the state-of-the-art approach Opiner, we found the improvement of RoBERTa varies from different aspects. 
For instance, in terms of the F1 score, RoBERTa achieves the greatest improvement on the aspect of Community by 70.6\%. However, for the aspect Usability, RoBERTa reaches the slightest improvement of 12.7\%.
In addition, we found that the improvements are significant, especially for the aspects with a small number of data instances.
For example, RoBERTa outperforms Opiner the most on the aspect Community, which only corresponds to 2.1\% instances of the whole dataset (see Table~\ref{tab:definitionofaspect}).
The aspect Usability contains 15 times more labeled instances than the aspect Community, while RoBERTa only achieves 12.7\% improvement over Opiner.
It indicates that PTM approaches are more capable of extracting semantic information than Opiner, especially when only a small number of data instances are available in the training set.
We discuss more about the rationale by conducting an error analysis in Section~\ref{sec:qualitative}.
We also observed that the best and the worst PTM approach varies in different API review aspects.
For example, XLNet achieves the best performance on the aspect Performance while it is the worst on the aspect Portability.

Compared with the prior work~\cite{zhang2020sentiment} which utilizes PTMs for sentiment analysis on software data, we have the same finding that PTMs outperform existing SOTA approaches by a large margin (ranges from 6.5\% to 35.6\% in~\cite{zhang2020sentiment} and from 21\% to 30.2\% in this work).
It indicates that PTMs have a great potential for boosting various types of SE tasks.
Furthermore, RoBERTa is also the overall winner in both \cite{zhang2020sentiment} and our work, which demonstrates that RoBERTa equips better generalization ability in SE domain tasks than other considered PTMs.
Thus, it suggests that considering RoBERTa as the first attempt for the following works.
\begin{tcolorbox}
The four well-known pre-trained transformer-based models consistently outperform the state-of-the-art tools, Opiner, on all the 11 aspects. 
The best transformer-based approach varies in different aspects and the improvement (over Opiner) of every PTM model varies in different aspects.
The performance improvements achieved by the PTM model ranges from 28.9\% to 30.2\% in terms of average F1 score for all API review aspects.
\end{tcolorbox}


\noindent\textbf{RQ2}: Can domain-specific BERT (\ie{} BERTOverflow) achieve better performance than the standard BERT? If not, can it still outperform Opiner?

\noindent \textbf{Motivation} Recent studies \cite{lan2019albert,gururangan2020don,tabassum2020code} have proven that, for many tasks, there are two typical ways to boost the model performance: (1) train on a larger dataset and (2) train on a domain-specific dataset. However, these two means usually cannot be applied simultaneously as large-scale domain-specific data is often unavailable. Thus, in this work, we are interested in investigating which one plays a more important role for aspect-based API review classification. To achieve that, we compare the standard BERT with its variant named BERTOverflow~\cite{tabassum2020code}. BERTOverflow follows the same architecture of BERT but is trained on the domain-specific data~(i.e., Stack Overflow). 


\noindent \textbf{Results \& Analysis}
To answer this research question, we report the performance of BERTOverflow for aspect-based API review classification.
As BERTOverflow is trained with BERT structure, we compare its performance with base version BERT on the API review dataset.

In Table~\ref{tab:avgf1fortransformer}, we observed that for all the aspects, the standard BERT consistently outperforms BERTOverflow for all the evaluation metrics.
Moreover, we found the performance gap between the standard BERT and BERTOverflow differs for different API review aspects.
For the aspect Compatibility, BERT slightly outperforms BERTOverflow by 0.5\% in terms of F1 score. However, for the aspect General Features, the standard BERT outperforms BERTOverflow by 13.8\% in terms of F1. 
The potential reason could be because of the size of the training data varies. The size of training data of BERT is 3.3 billion words~(from Wikipedia dataset and BookCorpus dataset) while which of BERTOverflow is 152M sentences.
Considering an average sentence length of BookCorpus is 11 words, the sentences used to pre-train the standard BERT is 2 times more than BERTOverflow.

In addition, we found that BERTOverflow still outperforms Opiner on all the aspects in terms of F1 score.
However, we found that BERTOverflow performs worse than Opiner in terms of MCC and AUC by 72\% and 23.8\%, respectively.



\begin{tcolorbox}
Standard BERT outperforms BERTOverflow by 3.9\%, 77.9\%, 28.8\% in terms of average F1, MCC, and AUC.
Compared with Opiner, BERTOverflow performs better in terms of average F1 by 25.3\% but worse on MCC and AUC by 72.0\% and 23.8\%.

\end{tcolorbox}


\noindent\textbf{RQ3}: Can the BERT for imbalanced data (\ie{} CostSensBERT) achieve better performance than the standard BERT? If not, can it still outperform Opiner?

\noindent \textbf{Motivation}
As the transformer-based models evolved rapidly, variants of PTMs emerged to solve specific tasks. CostSensBERT is one of them which is designed to deal with the data imbalance problem.
In the original paper of CostSensBERT~\cite{madabushi2019cost}, CostSensBERT achieves a better performance than BERT on the task of SE-specific named entity recognition on \so{} data.
In this study, we are interested in exploring the applicability of CostSensBERT to aspect-based API review classification.

\noindent \textbf{Results \& Analysis}
As shown in Table~\ref{tab:transformerperformance}, we observed that the performance of CostSensBERT is worse than BERT in terms of F1.
In particular, BERT outperforms CostSensBERT by 7.5\% in terms of F1.
However, CostSensBERT achieves better performance than BERT in terms of MCC by 8.89\%.
According to the definition of MCC mentioned in Section~\ref{sec:eval_metrics}, a high MCC can only be achieved when the predicted results are promising on all of the four confusion matrix categories (true positives, false negatives, true negatives, and false positives).
Therefore, compared with F1, MCC can shed a different light on the prediction ability of the minority class.
Our experimental results demonstrate that CostSensBERT improves the prediction ability of the minority class but hurts the prediction ability of majority samples to some extent.
In Table~\ref{tab:avgf1fortransformer}, compared with Opiner, CostSensBERT outperforms Opiner by 21.0\% in terms of average F1 score.
From Table V, we can find that CostSensBERT outperforms Opiner on all API review aspects.

Furthermore, the experimental results demonstrate the advantage of  CostSensBERT in handling the imbalanced dataset.
For instance, Usability and General Features are the two aspects with the most number of instances.
CostSensBERT outperforms BERT on aspect Usability by 2.7\% and 0.6\% in terms of AUC and MCC, respectively.
However, standard Bert outperforms CostSensBERT on the Usability aspect in terms of F1 by 3.2\%.
The same observation applies to the other aspects with a large number of instances, like General Features.
On the other hand, for the aspects with a small number of instances like Portability, CostSensBERT achieves better performance than BERT by 11.8\% and 4.5\% on AUC and MCC, respectively.
This confirms our finding that the CostSensBERT does help to handle imbalanced datasets by enhancing the ability to identify minority class samples, but at the same time, more majority class instances are mistakenly identified.
Thus, our results indicate that CostSensBERT may still be worthy of being considered.
Also, a model which leverages the advantages of standard BERT and CostSensBERT and avoids their disadvantages could be promising. And we leave it as our future work.

\begin{tcolorbox}
Standard BERT outperforms CostSensBERT in terms of average F1 by 7.5\%, but CostSensBERT performs better on average MCC and AUC by 8.9\% and 3.7\%, respectively. 
Moreover, CostSensBERT still outperforms Opiner by 21.0\%, 38.3\%, 10.9\% in terms of average F1, MCC and AUC.
\end{tcolorbox}



\section{Discussion}

\subsection{Error Analysis}\label{sec:qualitative}



We conduct an error analysis by (1) randomly sampling the predicted result produced by the considered approaches, (2) drawing Venn diagrams to interpret the relationship of the correctly predicted instances between the best performing PTM and Opiner.

Compared with Opiner, we observed that unseen features is one primary reason that all the considered PTMs achieve better performance.
During the prediction stage, Opiner is unable to capture the semantic meaning of the words when they do not exist in the vocabulary build on the training data.
However, benefiting from the pre-training on the huge amount of training data, PTMs can still capture partial semantic meaning.
Take an instance in our dataset as an example, the sentence  ``\emph{The javabean getters/setters have ... and does not slowdown...}'' is labeled as Performance. Opiner misclassifies this sentence while all PTMs predict the label correctly. We further found that the keyword ``\emph{slowdown}'' does not exist in the training dataset.
The example further proves that Opiner suffers from out-of-vocabulary issue~\cite{moon2020patchbert,hartmann2014large}.
At the same time, it also demonstrates the key advantage of pre-trained models.
\begin{figure}
    \centering
    \includegraphics[width=\linewidth]{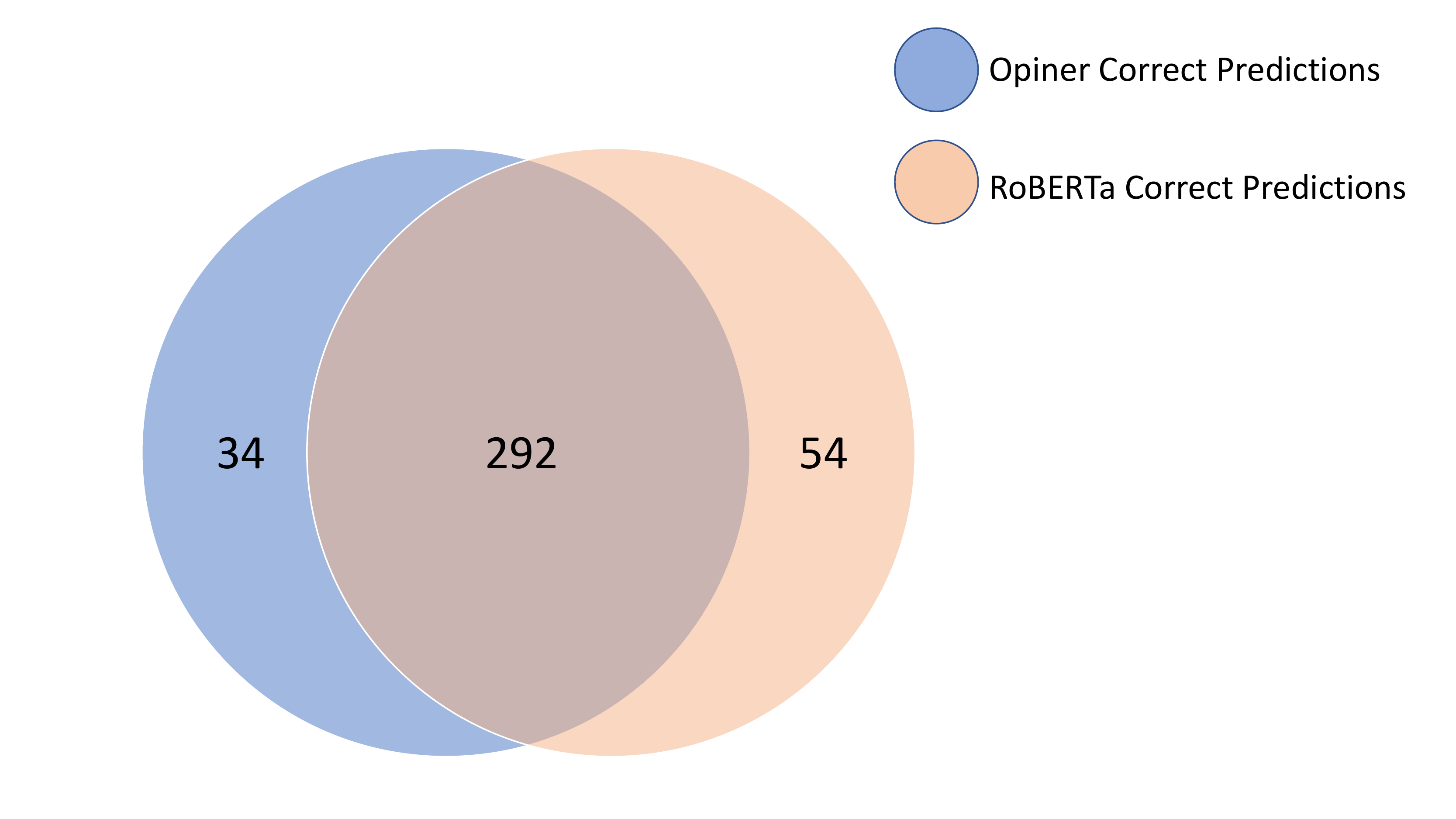}
    \vspace{-8mm}
    \caption{Venn Diagram of Aspect Usability}
    \label{fig:usability}
    \vspace{-6mm}
\end{figure}

\begin{figure}
    \centering
    \includegraphics[width=\linewidth]{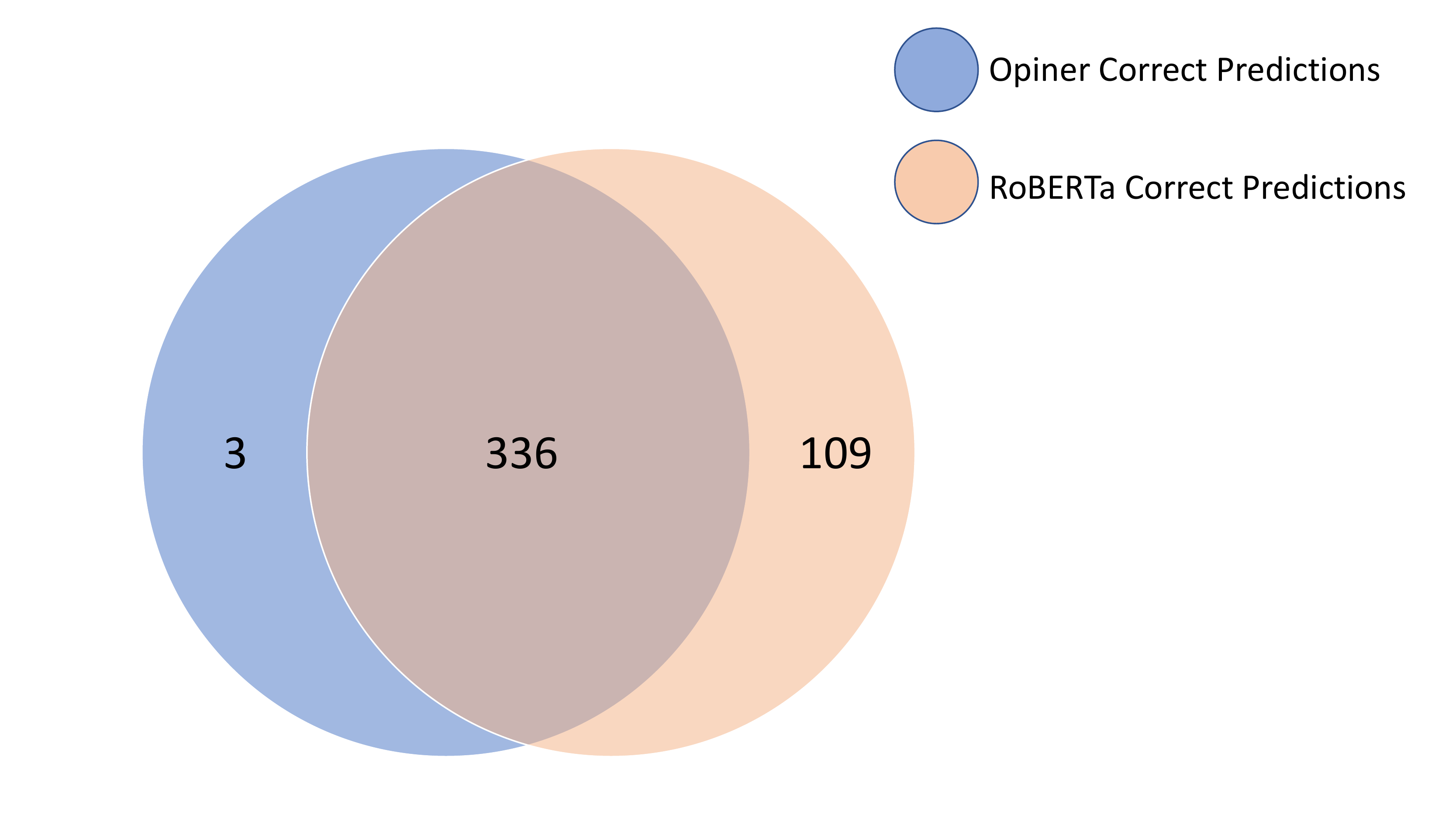}
    \vspace{-8mm}
    \caption{Venn Diagram of Aspect Portability}
    \label{fig:portability}
    \vspace{-6mm}
\end{figure}

Furthermore, we compare the best performing pre-trained transformer-based model, RoBERTA and Opiner on two selected API review aspects, Usability and Portability.
Usability and Portability are one of the aspects with the largest and smallest number of data instances, respectively.
The number of data instances labeled as Usability (1,437) is 15 times bigger than which of Portability (70).
In total, we have 448 instances in the test dataset.

We observed that some instances where only Opiner can produce the correct labels.
There are also cases where only RoBERTa can produce correct labels.
Figures \ref{fig:usability} and \ref{fig:portability} demonstrate the number of correct predictions produced by RoBERTa and Opiner in the two aspects Usability and Portability. For the aspect of Usability, RoBERTa and Opiner correctly predict 346 (77.2\%) and 326 (67.9\%) instances, respectively.
In addition, for 336 instances, both methods make the correct predictions.
For the API review aspect of portability, RoBERTa and Opiner correctly predict 445 (99.3\%) and 339 (75.6\%) instances, respectively.
In addition, for 292 instances, both methods make the correct predictions.
Thus we found that although Opiner performs worse than RoBERTa overall, it still performs better in some instances (i.e., 34 instances in Usability aspect, 3 instances in Portability aspect).
There is a potential to combine the two approaches to boost performance further. We leave this as future work.

\subsection{Parameter Settings}

Parameter settings play a critical role in determining the performance of deep learning models including PTMs.
In this section, we describe our observations on the best parameter setting of considered PTMs.
By this means, we aim to shed light on how to effectively adapt these PTMs for future research on aspect-based API review classification. Our findings may also possibly translate to other tasks.

In our experiments, we ran 10-fold cross-validation for every approach on all eleven aspects. In each fold, we fine-tune six PTMs twelve times based on different combinations of hyper-parameters described in Table~\ref{tab:modelparameter}.
Overall, we implement the fine-tuning process about 7,920 times ($10\times11\times6\times12$).
Moreover, as the dataset is imbalanced, our experiment covers different scales of classification tasks for the aspects (e.g., the number of instances in the Usability aspect is about 29 times of the Legal aspect), two sampling strategies, and 11 semantic labels (\ie{} aspects).
We summarize our observations and findings on parameter settings as below:

\noindent\textbf{1) Undersampling boosts the performance of traditional machine learning models but not for PTMs in this task.} We observed that in all the best experimental settings of Opiner, applying undersampling strategy (\ie{} with undersampling in Section~\ref{sec:sampling}) consistently achieves the best performance.
    For all the PTMs, we observed the opposite that undersampling does not boost and even slightly hurts the performance.

\noindent\textbf{2) Set the batch size to 32 produces better results than set it to 16 for most of the aspects.}
    We found that for all the considered PTMs, there is no golden combination of parameters that can achieve the best performance on all the aspects.
    One potential reason could be the diversity of the data in terms of the distribution and semantics across different aspects.
    However, we found that the batch size of 32 is preferable in most of the aspects thus we recommended future works to use the batch size as their first attempts.

\subsection{Threats to Validity}\label{sec:threats}
Threat to \textbf{internal validity} relates to errors and biases in our experiments.
To mitigate the threat, we employ a standard 10-fold cross-validation to avoid the bias that might be introduced to the results by the leave-one-out test dataset.
Also, we follow a standard hyper-parameter searching procedure on all PTMs\cite{devlin2019bert}.
As mentioned in Section~\ref{sec:implementations}, we re-implemented Opiner as its replication package is not online available.
We followed three steps to ensure the correctness of our replication: (1) we carefully followed all the technical details described in the original paper, (2) double check with the original authors about unclear information.
Due to possible differences in the random data shuffling in the data pre-processing stage,
our performance in all API review aspects is slightly higher than the performance of Opiner reported in the original paper\cite{uddin2019automatic}.
To replicate two specific variants of BERT (\ie{} BERTOverflow and CostSensBERT), we reused the replication package released by the original authors.
To implement the four popular PTMs, we utilized a widely-used deep learning library Hugging Face Transformer.
Considering above, we believe the threat is minimal.
Moreover, we will also release a replication package which includes both data and code
to facilitate validation and extension. In addition, one of the threats to internal validity refers to the quality of datasets labels. We inherit this threat as we use a publicly accessible dataset released from previous work.
Threats to \textbf{external validity} related to the generalizability of our results.
In contrast with the prior works which only consider mainstream PTMs~\cite{zhang2020sentiment}, we also consider variants of PTMs which might be potentially suitable considering the characteristics of the task.
For fair comparison, we reuse the same data that was released by Uddin et al. and used to evaluate Opiner~\cite{uddin2019automatic}. The findings that we have here may be different when additional PTMs and data are considered. We plan to consider more PTMs and a larger dataset than the one collected by Uddin et al. \cite{uddin2019automatic} in the future.
The threat to the \textbf{construct validity} of our work related to propriety of our evaluation metrics.
We follow the same evaluation metrics used in the prior works on the same tasks~\cite{uddin2017mining,uddin2019automatic}.
Moreover, the used evaluation metrics (\ie{} Precision, Recall, F1 score, MCC, and AUC) are widely used in the SE fields~\cite{xu2018prediction,xu2016predicting,xu2021post2vec}.


\section{Related Work}
 
 \subsection{Pre-trained transformer models for SE tasks}
Recently, pre-trained transformer models are receiving great attention for several domains including software engineering.
Zhang et al. fine-tuned BERT, RoBERTa, XLNet, ALBERT for sentiment analysis for software engineering (SA4SE) and compared them with five existing SA4SE tools on six datasets \cite{zhang2020sentiment}. They found that the PTMs outperform the traditional ones ranges from 6.5 to 35.6\% in terms of F1. 
Biswas et al. used BERT for sentiment analysis of software artifacts with significant improvement over the state-of-the-art \cite{biswas2020achieving}.
Khan et al. used BERT to detect different API documentation issues that outperformed all other models \cite{khan2021automatic}.
Moreover, Lin et al. utilize BERT and transfer learning to generate trace links between software text and source code \cite{lin2021traceability}.

Several transformer models can further be trained to learn code as well as contextual word representations and can be fine-tuned for different downstream SE tasks such as natural language code search, automated program repair, automated testing, code summarization. Tabassum et al. developed BERTOverflow, ELMoVerflow, GloVerflow for SE tasks by training the corresponding original versions (i.e. BERT, ELMo, GloVe) on 152 million Stack Overflow sentences \cite{tabassum2020code}. Based on the proposed models, they presented SoftNER for code and named entity recognition on Stack Overflow data.
Mosel et al. developed seBERT by pre-training it with 204.4 GB data coming from Stack Overflow, GitHub, and Jira \cite{von2021validity}. They found that seBERT can achieve promising results on multiple tasks, \eg{} quality improving commit identification and issue type prediction.
Feng et al. presented CodeBERT, a bimodal model for both natural and programming language which is pre-trained on a large dataset containing 2.1M bimodal data (i.e. code and corresponding documentation) and 6.4M unimodal data (i.e. only code) across six programming languages (Python, Java, JavaScript, PHP, Ruby, and Go) \cite{feng2020codebert}.
 
\subsection{API Review Analysis}
With rapid growth of APIs and their impact on software development process, developers often face difficulties to choose the right API. Hence, developers seek advice and reviews about different APIs from other developers in various online forums. 
However, the vast number of API reviews is hard to follow which poses a new challenge for the developers. Hence, several studies have focused on the analysis of such online developers discussion.
Hou et al.~\cite{hou2011obstacles} identified several categories of obstacles in using APIs and discussed about how to overcome those obstacles. 
Rosen et al. analyzed over 13M Stack Overflow posts by mobile developers which revealed various concerns such as various app distribution, mobile APIs, data management \cite{rosen2016mobile}. 
The automatic mining of API insights from such review discussion has also gotten a lot of attention.
Treude et al. detected insightful API related sentences from Stack Overflow discussion using ML techniques and used them to enhance API documentation \cite{treude2016augmenting}. 
Besides, Zhang et al. \cite{zhang2013extracting} extracted the problematic API features from online discussion by using sentiment analysis.
Being motivated by the presence of various API aspects (e.g., performance, usability) in their Stack Overflow case study, Uddin et al. developed a set of ML classifiers to detect them \cite{uddin2017mining, uddin2019automatic}. Their survey involving 178 developers shows that developers need API reviews and seek for automated tool support to assess them \cite{uddin2021understanding}. 
Ahasanuzzaman et al. focus on identifying API issue-related posts in \so{}.
To address the problem, they develop a supervised learning approach CAPS, which based on five different dimensions and conditional random field (CRF) technique~\cite{ahasanuzzaman2018classifying}.
In the following work, Ahasanuzzaman et al. further extend their approach by proposing more features~\cite{ahasanuzzaman2020caps}.

\section{Conclusion and Future Work}
In this study, we conducted an empirical evaluation on the performance of PTMs (pre-trained transformer-based models) and the state-of-the-art machine learning-based approach for API review classification. We are the first to shed a light of the effectiveness among variants of PTMs (i.e., domain-specific, designed for imbalanced data). 
Our empirical study includes six PTMs, i.e., BERT, RoBERTa, ALBERT, XLNET, BERTOverflow, and CostSensBERT.
They cover not only the state-of-the-art PTMs but also PTMs that are designed considering the characteristics of the data that we have (e.g., SE domain-specific data, imbalanced label distribution).
Our experimental result shows that all the PTMs consistently outperform the state-of-the-art approaches.
More importantly, we found that two potential suitable variants of BERT do not perform better than the standard BERT in our task.
However, particularly, CostSensBERT sheds a light on its ability on identifying the minority class that it achieves higher MCC than the standard BERT. Hence, we suggest that CostSensBERT is still worth trying for SE tasks with imbalanced data.

Overall, PTMs are more applicable than the existing state-of-the-art approach in API review classification task. We require a more in-depth look at the relationship among PTMs different from data domain and design purposes.
We release the replication package \footnote{\url{https://github.com/soarsmu/PTM4SE}}.
In the future, we plan to explore the following directions for further boosting aspect-based API review classification: (1) investigate more potentially suitable PTM variants, and (2) design an ensemble PTM-based model to leverage the advantage of RoBERTa and Opiner and avoid their shortcomings at the same time.


\section*{ACKNOWLEDGMENT}
\vspace{-1mm}
This research / project is supported by the Ministry of Education, Singapore, under its Academic Research Fund Tier 2 (Award No.: MOE2019-T2-1-193). Any opinions, findings and conclusions or recommendations expressed in this material are those of the author(s) and do not reflect the views of the Ministry of Education, Singapore. This research / project is also supported by Natural Sciences and Engineering Research Council of Canada (NSERC), University of Calgary, and Alberta Innovates.

\footnotesize
\bibliographystyle{IEEEtran}
\vspace{-2mm}
\bibliography{references}

\begin{thebibliography}{10}
\providecommand{\url}[1]{#1}
\csname url@samestyle\endcsname
\providecommand{\newblock}{\relax}
\providecommand{\bibinfo}[2]{#2}
\providecommand{\BIBentrySTDinterwordspacing}{\spaceskip=0pt\relax}
\providecommand{\BIBentryALTinterwordstretchfactor}{4}
\providecommand{\BIBentryALTinterwordspacing}{\spaceskip=\fontdimen2\font plus
\BIBentryALTinterwordstretchfactor\fontdimen3\font minus
  \fontdimen4\font\relax}
\providecommand{\BIBforeignlanguage}[2]{{%
\expandafter\ifx\csname l@#1\endcsname\relax
\typeout{** WARNING: IEEEtran.bst: No hyphenation pattern has been}%
\typeout{** loaded for the language `#1'. Using the pattern for}%
\typeout{** the default language instead.}%
\else
\language=\csname l@#1\endcsname
\fi
#2}}
\providecommand{\BIBdecl}{\relax}
\BIBdecl

\bibitem{ahasanuzzaman2018classifying}
M.~Ahasanuzzaman, M.~Asaduzzaman, C.~K. Roy, and K.~A. Schneider, ``Classifying
  stack overflow posts on api issues,'' in \emph{2018 IEEE 25th international
  conference on software analysis, evolution and reengineering (SANER)}.\hskip
  1em plus 0.5em minus 0.4em\relax IEEE, 2018, pp. 244--254.

\bibitem{uddin2021understanding}
G.~Uddin, O.~Baysal, L.~Guerrouj, and F.~Khomh, ``Understanding how and why
  developers seek and analyze api-related opinions,'' \emph{IEEE Transactions
  on Software Engineering}, vol.~47, no.~04, pp. 694--735, 2021.

\bibitem{ahasanuzzaman2020caps}
M.~Ahasanuzzaman, M.~Asaduzzaman, C.~K. Roy, and K.~A. Schneider, ``Caps: a
  supervised technique for classifying stack overflow posts concerning api
  issues,'' \emph{Empirical Software Engineering}, vol.~25, no.~2, pp.
  1493--1532, 2020.

\bibitem{uddin2015api}
G.~Uddin and M.~P. Robillard, ``How api documentation fails,'' \emph{Ieee
  software}, vol.~32, no.~4, pp. 68--75, 2015.

\bibitem{uddin2017mining}
G.~Uddin and F.~Khomh, ``Mining api aspects in api reviews,'' in
  \emph{Technical Report}, 2017.

\bibitem{uddin2017automatic}
{G. Uddin and F. Khomh}, ``Automatic summarization of api reviews,'' in
  \emph{2017 32nd IEEE/ACM International Conference on Automated Software
  Engineering (ASE)}.\hskip 1em plus 0.5em minus 0.4em\relax IEEE, 2017, pp.
  159--170.

\bibitem{lin2019pattern}
B.~Lin, F.~Zampetti, G.~Bavota, M.~Di~Penta, and M.~Lanza, ``Pattern-based
  mining of opinions in q\&a websites,'' in \emph{2019 IEEE/ACM 41st
  International Conference on Software Engineering (ICSE)}.\hskip 1em plus
  0.5em minus 0.4em\relax IEEE, 2019, pp. 548--559.

\bibitem{uddin2019automatic}
G.~Uddin and F.~Khomh, \, ``Automatic mining of opinions expressed about apis
  in stack overflow,'' \emph{IEEE Transactions on Software Engineering},
  vol.~47, no.~3, pp. 522--559, 2021.

\bibitem{zhang2020sentiment}
T.~Zhang, B.~Xu, F.~Thung, S.~A. Haryono, D.~Lo, and L.~Jiang, ``Sentiment
  analysis for software engineering: How far can pre-trained transformer models
  go?'' in \emph{2020 IEEE International Conference on Software Maintenance and
  Evolution (ICSME)}.\hskip 1em plus 0.5em minus 0.4em\relax IEEE, 2020, pp.
  70--80.

\bibitem{zhou2019lancer}
S.~Zhou, B.~Shen, and H.~Zhong, ``Lancer: Your code tell me what you need,'' in
  \emph{2019 34th IEEE/ACM International Conference on Automated Software
  Engineering (ASE)}.\hskip 1em plus 0.5em minus 0.4em\relax IEEE, 2019, pp.
  1202--1205.

\bibitem{yu2020order}
Z.~Yu, R.~Cao, Q.~Tang, S.~Nie, J.~Huang, and S.~Wu, ``Order matters:
  Semantic-aware neural networks for binary code similarity detection,'' in
  \emph{Proceedings of the AAAI Conference on Artificial Intelligence},
  vol.~34, no.~01, 2020, pp. 1145--1152.

\bibitem{wang2020fret}
R.~Wang, H.~Zhang, G.~Lu, L.~Lyu, and C.~Lyu, ``Fret: Functional reinforced
  transformer with bert for code summarization,'' \emph{IEEE Access}, vol.~8,
  pp. 135\,591--135\,604, 2020.

\bibitem{devlin2019bert}
J.~Devlin, M.-W. Chang, K.~Lee, and K.~Toutanova, ``Bert: Pre-training of deep
  bidirectional transformers for language understanding,'' in \emph{Proceedings
  of the 2019 Conference of the North American Chapter of the Association for
  Computational Linguistics: Human Language Technologies, Volume 1 (Long and
  Short Papers)}, 2019, pp. 4171--4186.

\bibitem{liu2019roberta}
Y.~Liu, M.~Ott, N.~Goyal, J.~Du, M.~Joshi, D.~Chen, O.~Levy, M.~Lewis,
  L.~Zettlemoyer, and V.~Stoyanov, ``Roberta: A robustly optimized bert
  pretraining approach,'' \emph{arXiv preprint arXiv:1907.11692}, 2019.

\bibitem{lan2019albert}
Z.~Lan, M.~Chen, S.~Goodman, K.~Gimpel, P.~Sharma, and R.~Soricut, ``Albert: A
  lite bert for self-supervised learning of language representations,''
  \emph{arXiv preprint arXiv:1909.11942}, 2019.

\bibitem{yang2019xlnet}
Z.~Yang, Z.~Dai, Y.~Yang, J.~Carbonell, R.~R. Salakhutdinov, and Q.~V. Le,
  ``Xlnet: Generalized autoregressive pretraining for language understanding,''
  \emph{Advances in neural information processing systems}, vol.~32, 2019.

\bibitem{gururangan2020don}
S.~Gururangan, A.~Marasovi{\'c}, S.~Swayamdipta, K.~Lo, I.~Beltagy, D.~Downey,
  and N.~A. Smith, ``Don't stop pretraining: adapt language models to domains
  and tasks,'' \emph{arXiv preprint arXiv:2004.10964}, 2020.

\bibitem{tabassum2020code}
J.~Tabassum, M.~Maddela, W.~Xu, and A.~Ritter, ``Code and named entity
  recognition in stackoverflow,'' in \emph{Proceedings of the 58th Annual
  Meeting of the Association for Computational Linguistics}, 2020, pp.
  4913--4926.

\bibitem{madabushi2019cost}
H.~T. Madabushi, E.~Kochkina, and M.~Castelle, ``Cost-sensitive bert for
  generalisable sentence classification with imbalanced data,''
  \emph{EMNLP-IJCNLP 2019}, p. 125, 2019.

\bibitem{yang2021deepscc}
G.~Yang, Y.~Zhou, C.~Yu, and X.~Chen, ``Deepscc: Source code classification
  based on fine-tuned roberta,'' 2021.

\bibitem{dai2019transformer}
Z.~Dai, Z.~Yang, Y.~Yang, J.~Carbonell, Q.~V. Le, and R.~Salakhutdinov,
  ``Transformer-xl: Attentive language models beyond a fixed-length context,''
  \emph{arXiv preprint arXiv:1901.02860}, 2019.

\bibitem{uddin2019understanding}
G.~Uddin, O.~Baysal, L.~Guerrouj, and F.~Khomh, ``Understanding how and why
  developers seek and analyze api-related opinions,'' \emph{IEEE Transactions
  on Software Engineering}, 2019.

\bibitem{loshchilov2017decoupled}
I.~Loshchilov and F.~Hutter, ``Decoupled weight decay regularization,''
  \emph{arXiv preprint arXiv:1711.05101}, 2017.

\bibitem{biswas2020achieving}
E.~Biswas, M.~E. Karabulut, L.~Pollock, and K.~Vijay-Shanker, ``Achieving
  reliable sentiment analysis in the software engineering domain using bert,''
  in \emph{2020 IEEE International Conference on Software Maintenance and
  Evolution (ICSME)}.\hskip 1em plus 0.5em minus 0.4em\relax IEEE, 2020, pp.
  162--173.

\bibitem{khan2021automatic}
J.~Y. Khan, M.~T.~I. Khondaker, G.~Uddin, and A.~Iqbal, ``Automatic detection
  of five api documentation smells: Practitioners’ perspectives,'' in
  \emph{2021 IEEE International Conference on Software Analysis, Evolution and
  Reengineering (SANER)}.\hskip 1em plus 0.5em minus 0.4em\relax IEEE, 2021,
  pp. 318--329.

\bibitem{hand2001simple}
D.~J. Hand and R.~J. Till, ``A simple generalisation of the area under the roc
  curve for multiple class classification problems,'' \emph{Machine learning},
  vol.~45, no.~2, pp. 171--186, 2001.

\bibitem{huang2005using}
J.~Huang and C.~X. Ling, ``Using auc and accuracy in evaluating learning
  algorithms,'' \emph{IEEE Transactions on knowledge and Data Engineering},
  vol.~17, no.~3, pp. 299--310, 2005.

\bibitem{moon2020patchbert}
S.~Moon and N.~Okazaki, ``Patchbert: Just-in-time, out-of-vocabulary
  patching,'' in \emph{Proceedings of the 2020 Conference on Empirical Methods
  in Natural Language Processing (EMNLP)}, 2020, pp. 7846--7852.

\bibitem{hartmann2014large}
N.~Hartmann, L.~Avan{\c{c}}o, P.~Balage~Filho, M.~S. Duran, M.~D. G.~V. Nunes,
  T.~Pardo, and S.~Alu{\'\i}sio, ``A large corpus of product reviews in
  portuguese: Tackling out-of-vocabulary words,'' in \emph{Proceedings of the
  Ninth International Conference on Language Resources and Evaluation
  (LREC'14)}, 2014, pp. 3865--3871.

\bibitem{xu2018prediction}
B.~Xu, A.~Shirani, D.~Lo, and M.~A. Alipour, ``Prediction of relatedness in
  stack overflow: deep learning vs. svm: a reproducibility study,'' in
  \emph{Proceedings of the 12th ACM/IEEE International Symposium on Empirical
  Software Engineering and Measurement}, 2018, pp. 1--10.

\bibitem{xu2016predicting}
B.~Xu, D.~Ye, Z.~Xing, X.~Xia, G.~Chen, and S.~Li, ``Predicting semantically
  linkable knowledge in developer online forums via convolutional neural
  network,'' in \emph{Proceedings of the 31st IEEE/ACM International Conference
  on Automated Software Engineering}.\hskip 1em plus 0.5em minus 0.4em\relax
  ACM, 2016, pp. 51--62.

\bibitem{xu2021post2vec}
B.~Xu, T.~Hoang, A.~Sharma, C.~Yang, X.~Xia, and D.~Lo, ``Post2vec: Learning
  distributed representations of stack overflow posts,'' \emph{IEEE
  Transactions on Software Engineering}, 2021.

\bibitem{lin2021traceability}
J.~Lin, Y.~Liu, Q.~Zeng, M.~Jiang, and J.~Cleland-Huang, ``Traceability
  transformed: Generating more accurate links with pre-trained bert models,''
  in \emph{2021 IEEE/ACM 43rd International Conference on Software Engineering
  (ICSE)}.\hskip 1em plus 0.5em minus 0.4em\relax IEEE, 2021, pp. 324--335.

\bibitem{von2021validity}
J.~von~der Mosel, A.~Trautsch, and S.~Herbold, ``On the validity of pre-trained
  transformers for natural language processing in the software engineering
  domain,'' \emph{arXiv preprint arXiv:2109.04738}, 2021.

\bibitem{feng2020codebert}
Z.~Feng, D.~Guo, D.~Tang, N.~Duan, X.~Feng, M.~Gong, L.~Shou, B.~Qin, T.~Liu,
  D.~Jiang \emph{et~al.}, ``Codebert: A pre-trained model for programming and
  natural languages,'' \emph{arXiv preprint arXiv:2002.08155}, 2020.

\bibitem{hou2011obstacles}
D.~Hou and L.~Li, ``Obstacles in using frameworks and apis: An exploratory
  study of programmers' newsgroup discussions,'' in \emph{2011 IEEE 19th
  International Conference on Program Comprehension}.\hskip 1em plus 0.5em
  minus 0.4em\relax IEEE, 2011, pp. 91--100.

\bibitem{rosen2016mobile}
C.~Rosen and E.~Shihab, ``What are mobile developers asking about? a large
  scale study using stack overflow,'' \emph{Empirical Software Engineering},
  vol.~21, no.~3, pp. 1192--1223, 2016.

\bibitem{treude2016augmenting}
C.~Treude and M.~P. Robillard, ``Augmenting api documentation with insights
  from stack overflow,'' in \emph{2016 IEEE/ACM 38th International Conference
  on Software Engineering (ICSE)}.\hskip 1em plus 0.5em minus 0.4em\relax IEEE,
  2016, pp. 392--403.

\bibitem{zhang2013extracting}
Y.~Zhang and D.~Hou, ``Extracting problematic api features from forum
  discussions,'' in \emph{2013 21st International Conference on Program
  Comprehension (ICPC)}.\hskip 1em plus 0.5em minus 0.4em\relax IEEE, 2013, pp.
  142--151.

\end{thebibliography}

\balance

\end{document}